# DeFi: Concepts and Ecosystem

Carlos J. Costa
ISEG, Universidade de Lisboa
Lisboa, Portugal
cjcosta@iseg.ulisboa.pt

*Abstract* - **This paper investigates the evolving landscape of decentralized finance (DeFi) by examining its foundational concepts, research trends, and ecosystem. A bibliometric analysis was conducted to identify thematic clusters and track the evolution of DeFi research. Additionally, a thematic review was performed to analyze the roles and interactions of key participants within the DeFi ecosystem, focusing on its opportunities and inherent risks. The bibliometric analysis identified a progression in research priorities, transitioning from an initial focus on technological innovation to addressing sustainability, environmental impacts, and regulatory challenges. Key thematic clusters include decentralization, smart contracts, tokenization, and sustainability concerns. The analysis of participants highlighted the roles of developers, liquidity providers, auditors, and regulators while identifying critical risks such as smart contract vulnerabilities, liquidity constraints, and regulatory uncertainties. The study underlines the transformative potential of DeFi to enhance financial inclusion and transparency while emphasizing the need for robust security frameworks and regulatory oversight to ensure long-term stability. This paper comprehensively explains the DeFi ecosystem by integrating bibliometric and thematic analyses. It offers valuable insights for researchers, practitioners, and policymakers, contributing to the ongoing discourse on the sustainable development and integration of DeFi into the global financial system.**

## I. Introduction

The development of blockchain has an impact in many fields, as referenced in the literature [1] [2] [3] [4], not only in cryptocurrencies[5] [6] or NFT. Decentralized finance (DeFi) has emerged as a transformative innovation in the financial sector, leveraging blockchain technology to offer decentralized alternatives to traditional financial services such as banking, lending, trading, and asset management. By eliminating intermediaries, DeFi systems aim to foster transparency, efficiency, and accessibility, potentially reshaping the global financial landscape. At its core, DeFi promotes financial inclusivity, enabling individuals worldwide to access essential financial services without reliance on traditional institutions. Furthermore, DeFi platforms empower users by granting greater control over their assets through self-custody and direct participation in financial markets. However, DeFi remains in its nascent stages and faces significant challenges. Technical vulnerabilities, such as flaws in smart contracts, economic risks linked to token volatility, and regulatory uncertainties, pose substantial barriers. Additionally, mainstream adoption is limited due to a lack of trust and understanding of decentralized systems. Addressing these challenges is critical to realizing DeFi's decentralized and equitable financial ecosystem promise.[7]

This paper aims to comprehensively analyze the DeFi ecosystem, identifying its core components, participants, and underlying dynamics. This paper has V sections. Section II introduces the DeFi ecosystem's main concepts, offering foundational knowledge for understanding its structure and operation. Section III details the method, including a bibliometric analysis and network-based approaches to explore the ecosystem and its participants. Section IV presents an overview of the DeFi ecosystem through a bibliometric analysis, mapping key themes, trends, and the evolution of research in the field. Section V examines the diverse participants in the DeFi ecosystem, highlighting their roles and interactions, including developers, liquidity providers, auditors, and regulators. Section VI explores the network connecting the actors in the DeFi ecosystem, uncovering relationships, collaborations, and the flow of influence within the decentralized financial landscape.

Summarizing, this paper offers a holistic view of the DeFi ecosystem, addressing its opportunities and challenges while contributing to ongoing discussions about its future development. This review serves as a resource for researchers, practitioners, and policymakers seeking to understand the potential and limitations of DeFi in shaping the future of global finance.

## II. Main Concepts

While DeFi presents potential for the financial landscape, addressing the inherent risks and developing robust security frameworks to foster trust and stability in these emerging systems is essential. Decentralized Finance (DeFi) refers to a financial ecosystem built on blockchain technology, aiming to provide financial services without traditional intermediaries like banks or brokers. Here are the main concepts of Defi: (1) Decentralization, (2) Smart Contracts, (3) Tokenization, (4) Decentralized Exchanges (DEXs), (5) Liquidity Pools, (6) Yield Farming and Staking, (7) Lending and Borrowing, (8) Governance, (9) Interoperability, (10) Risk and Security.

DeFi operates on decentralized platforms, typically using blockchain technology (e.g., Ethereum, Solana).[8] Transactions are validated by distributed nodes, ensuring transparency and security without centralized control.

Smart Contracts are Self-executing contracts with the terms of the agreement directly written in code [7]. Smart contracts automate transactions and enforce agreements without intermediaries, ensuring security and efficiency [9]. Use Cases include automated financial operations like lending, borrowing, and trading. For Example, a lending platform where loans are issued automatically when collateral conditions are met.

Tokenization means representing assets (e.g., currency, real estate) as tokens on a blockchain [10]. Tokenization allows for the representation of real-world assets on the blockchain, broadening investment opportunities and enhancing liquidity [9]. There are two types of tokens: tokens and stablecoins. Utility Tokens provide access to specific services (e.g., UNI for Uniswap governance). Stablecoins pegged to real-world assets like USD (e.g., USDT, DAI) to minimize volatility.

Decentralized Exchanges (DEXs) allow peer-to-peer trading of cryptocurrencies without intermediaries. The key mechanism are the AMMs (automated market makers) to facilitate trades through liquidity pools. DEXs enable trustless trading of cryptocurrencies, with AMMs like Uniswap or SushiSwap, facilitating liquidity through user-contributed pools [11] [12].

Liquidity Pools are pools of tokens locked in smart contracts to provide liquidity for decentralized exchanges and lending protocols. DEXs enable trustless trading of cryptocurrencies, with AMMs facilitating liquidity through user-contributed pools [11]. Benefits are earning fees or rewards for contributing liquidity and enabling seamless trades in DEXs.

Yield Farming means earning rewards (e.g., tokens) by providing liquidity or participating in lending protocols. Yield farming and staking allow users to earn passive income by locking their assets in DeFi protocols [13]. Staking is locking up tokens to secure a blockchain network or earn interest/rewards.[14]

The lending and borrowing mechanism corresponds to the borrower's deposit collateral to take loans, and the lender earns interest by supplying liquidity. Examples of platforms are Aave and Compound.[15]

In terms of governance, DeFi protocols often allow token holders to participate in decision-making (e.g., protocol upgrades).[16] Governance is often facilitated through decentralized autonomous organizations (DAOs), which utilize smart contracts for decision-making, enhancing transparency and reducing information asymmetries[8]. An example of a DAO is MakerDAO.[17]

Interoperability is also important in Defi. The purpose is to have different DeFi platforms and blockchains work together via cross-chain bridges or interoperable standards. The goal is to create a unified financial system without silos.[18]

Despite the advantages, DeFi faces risks such as smart contract vulnerabilities, liquidity challenges, and regulatory uncertainties [13]. Security measures are crucial to mitigate these risks and ensure the sustainability of DeFi systems. Amon risks, there are smart contract bugs[19], market volatility [20], and impermanent loss [21] in liquidity provision. In order to mitigate this, some measures may be taken, like code audits and decentralized insurance platforms.[22]

Among the main applications of DeFi are payments, derivatives, and savings. Payments tend to be fast and cheap specially in cross-border transactions. Derivatives consist of trading contracts based on asset values. Savings allow high-interest accounts via lending protocols.[23]

## III. Method

The study presented in this paper employed a multi-faceted methodological approach to analyze decentralized finance (DeFi) information, combining quantitative and qualitative analyses. The steps included mapping the evolution of key topics, deep analysis of abstracts, and identifying DeFi key actors and relationships.

The first step involves identifying the primary topics within DeFi research and tracking their evolution over time. This was achieved through the analysis of keywords and abstracts from selected papers. Keywords were extracted and analyzed for co-occurrences, revealing prominent themes and their interconnections. Abstracts provided a broader context, offering insights into the research focus and thematic shifts.

Abstracts of the selected papers were further scrutinized to uncover detailed information about the main areas of interest within DeFi. This step aimed to distill core concepts, trends, and emerging issues, ensuring a comprehensive understanding of the field.

The study identified the main entities involved in the DeFi ecosystem, such as researchers, institutions, and technologies. The analysis also explored relationships between these actors, shedding light on collaborations, influences, and network structures within the DeFi landscape.

## IV. DeFi Ecosystem: An Overview

A Scopus search using the query `TITLE-ABS-KEY ("decentralized finance")` identified 2,507 papers. The data was mapped using VOSviewer [24], focusing on keyword co-occurrence analysis. Initially, 14,458 unique keywords were identified. A minimum occurrence threshold of five was applied to refine the analysis, reducing the number of keywords to 970. The term "Decentralized Finance" was excluded to concentrate on related but distinct themes. This analysis revealed eight clusters: Investment, Cryptocurrencies, Decentralization, Performance, Green Economy, Consumption, Cybersecurity, and ESG (Environmental, Social, and Governance).

The analysis showed early research focused on technological aspects like blockchain and cryptographic mechanisms. Over time, the emphasis shifted towards broader environmental and social concerns, with sustainability becoming a key area of interest. Keywords related to green economy and ESG factors appeared more frequently in recent studies, reflecting the growing importance of environmental and social impacts within decentralized finance.

A second analysis was conducted on the same dataset, this time examining the abstracts and titles of the papers. Using binary counting, 46,568 unique words were identified. Applying a minimum occurrence threshold of ten reduced the terms to 1,264. The top 60% of the most relevant terms were selected for further exploration. This analysis uncovered four main clusters: Blockchain Technology, Pricing and Risks, Impact, and Decision-Making.

The overlay visualization from this analysis provided additional insights into the evolution of decentralized finance research. Early studies focused on technological frameworks and innovations, such as developing and optimizing blockchain systems. More recent works, however, have shifted their focus to the sustainability of decentralized finance, particularly in terms of its energy consumption and environmental impacts. This evolution underscores the increasing prioritization of sustainable practices within the decentralized finance ecosystem.

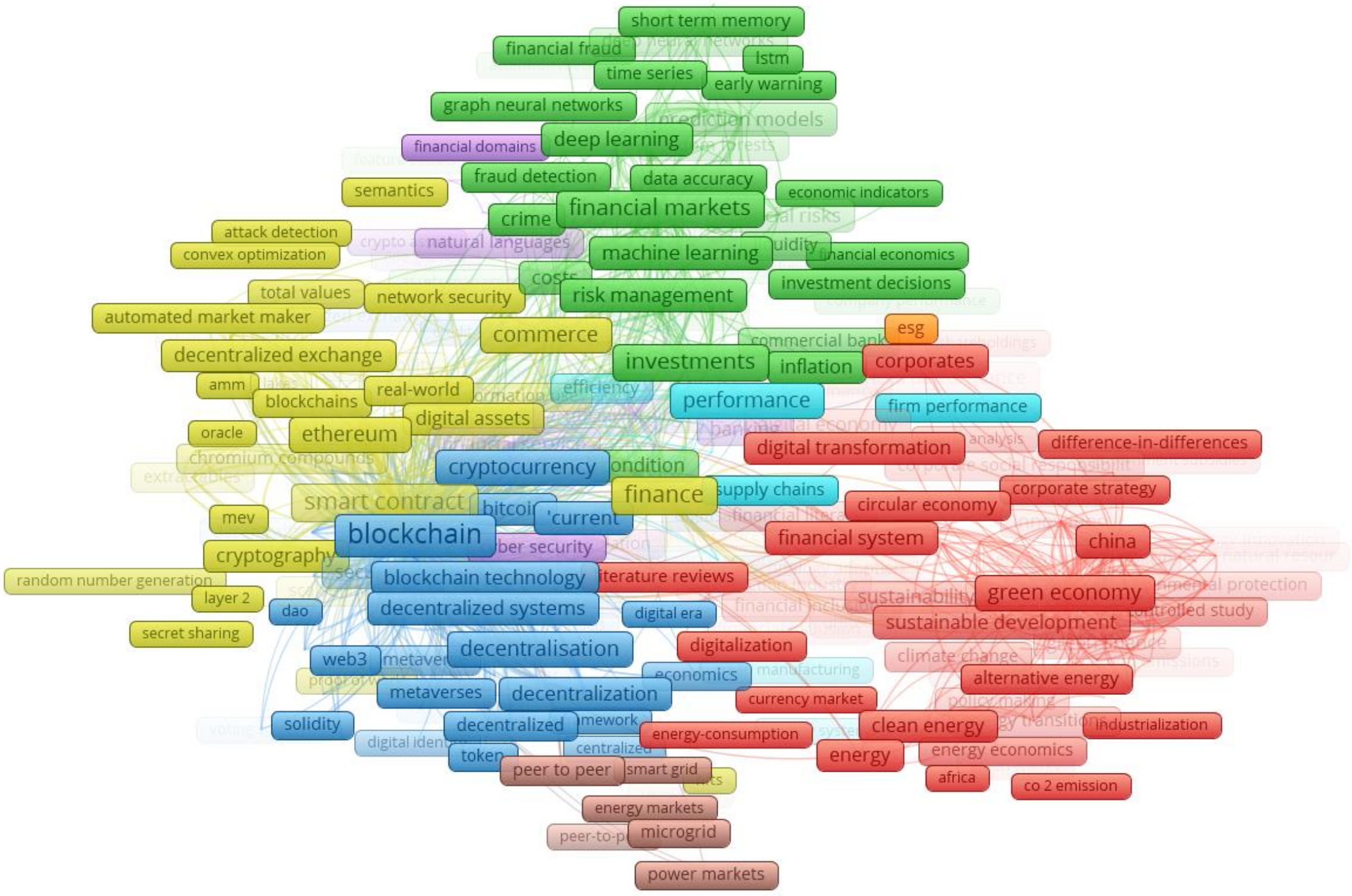

Fig.1 Network visualization of the keywords co-occurrences

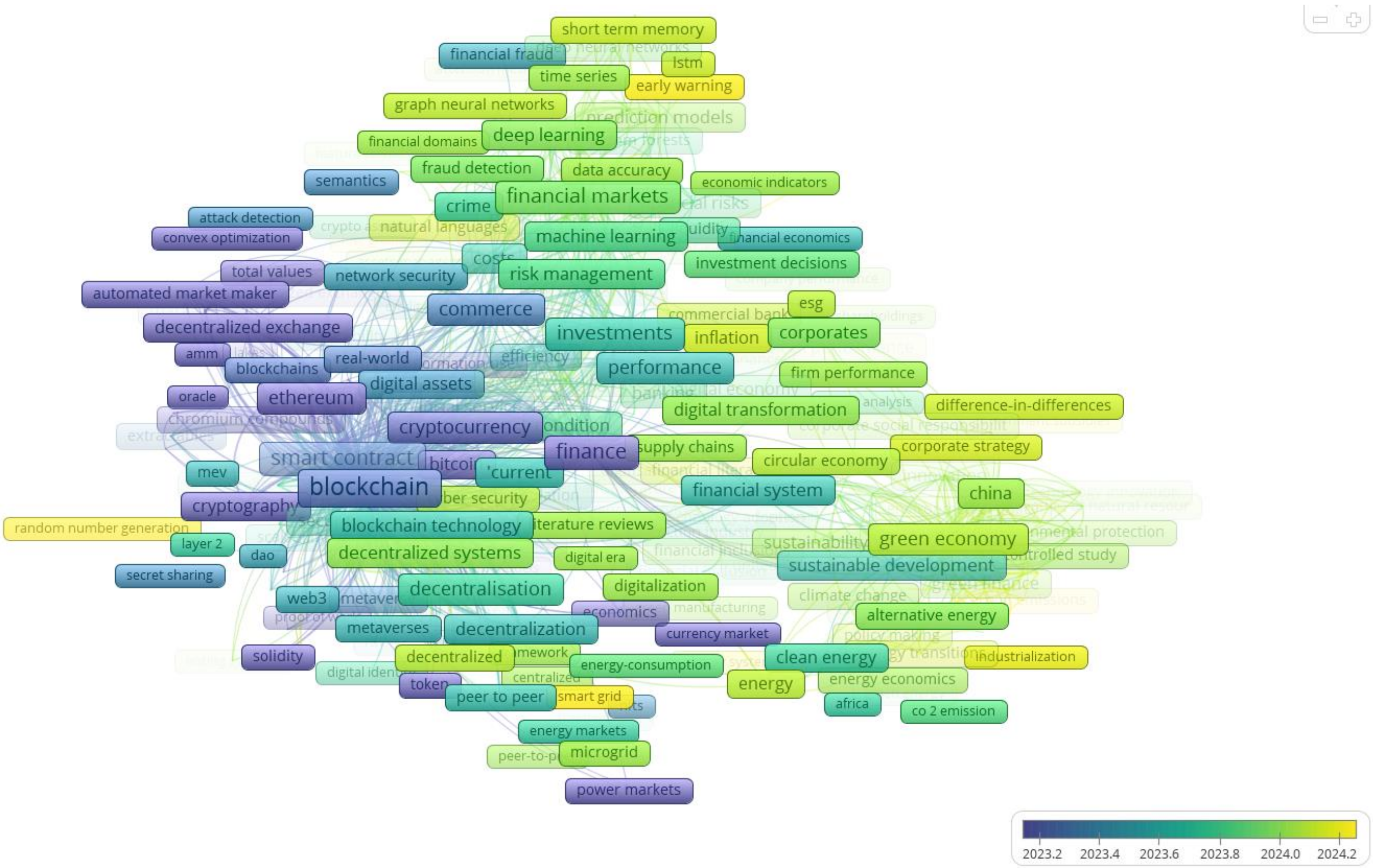

Fig. 2 Overlay visualization of the keywords co-occurrences

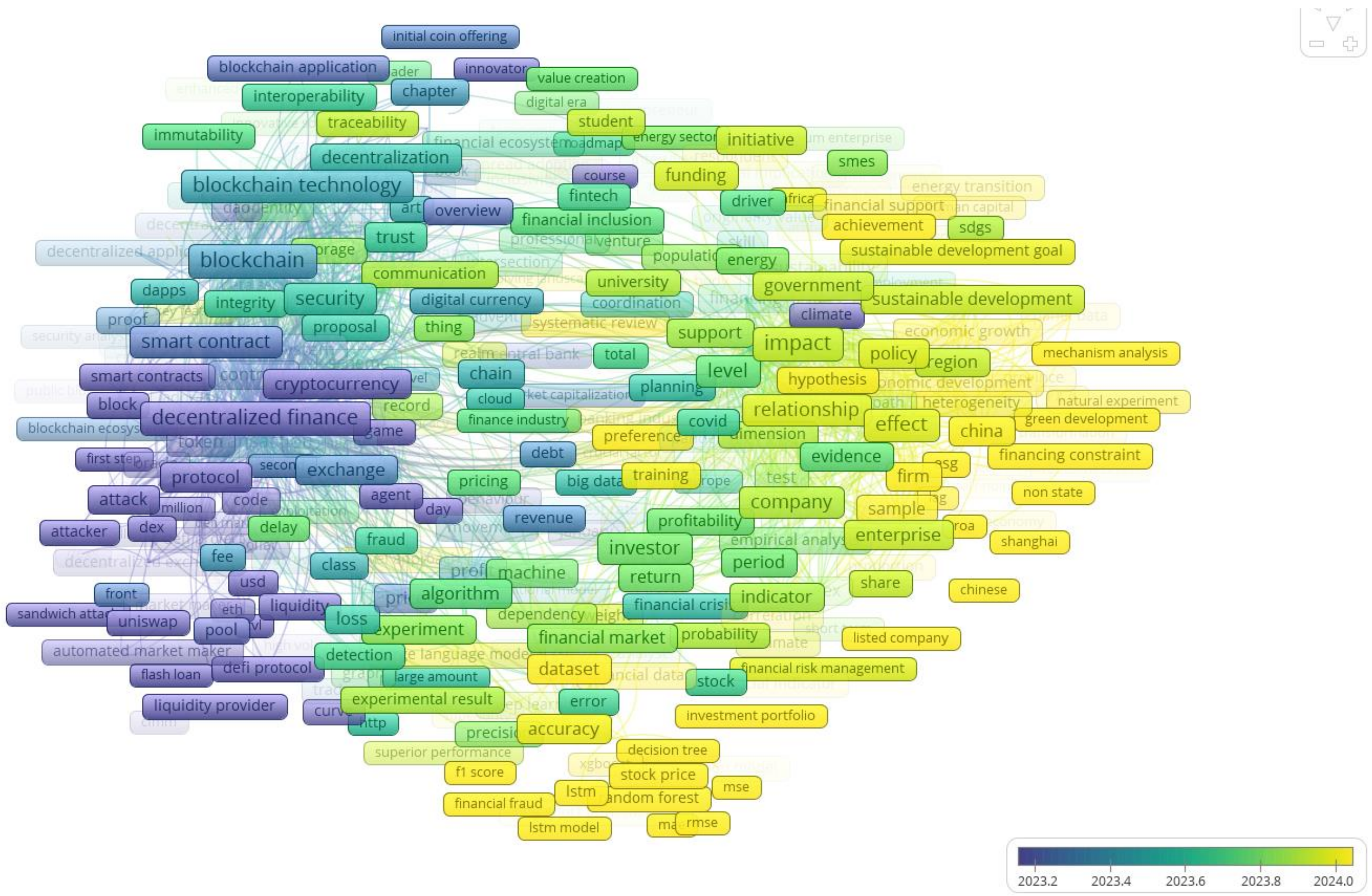

Fig.3   Network visualization of the terms in abstracts

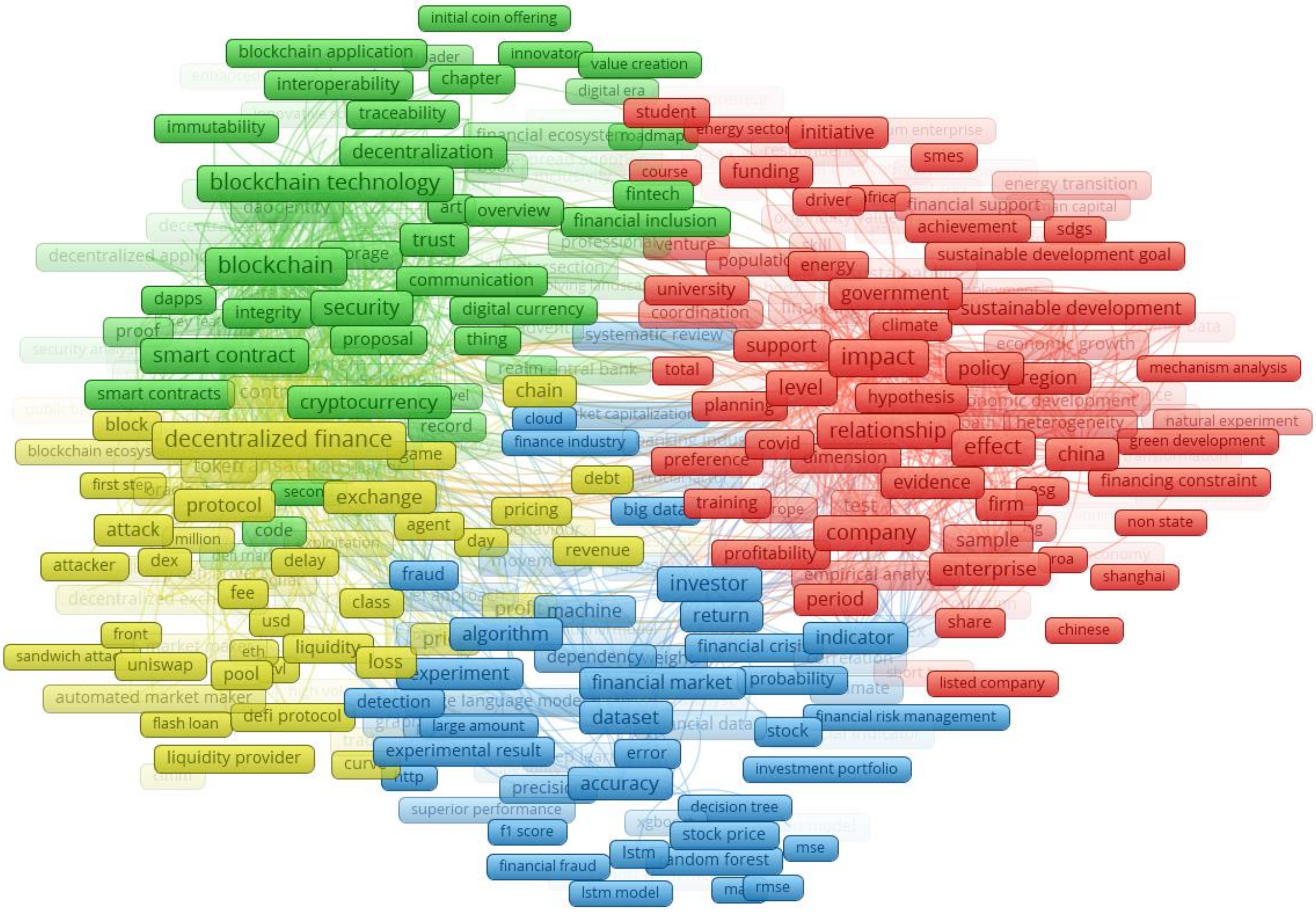

Fig. 4   Overlay visualization of terms the binary counting

## V. DEFI ECOSYSTEM: PARTICIPANTS

The Decentralized Finance (DeFi) ecosystem comprises diverse participants, each playing a crucial role in its functionality and growth. Each group contributes uniquely to the DeFi landscape, facilitating various financial services and governance structures.

In the DeFi ecosystem, participants play various roles, interacting with protocols, assets, and platforms to enable financial activities. Here are the main participants in DeFi:

1. Developers
2. Protocol Operators
3. Liquidity Providers (LPs)
4. Borrowers
5. Lenders
6. Traders
7. Investors and Speculators
8. Governance Participants
9. Oracles
10. Auditors and Security Firms
11. Decentralized Autonomous Organizations (DAOs)
12. End Users
13. Insurance Providers
14. Regulators

Developers [3] create and maintain DeFi protocols, ensuring functionality and security, while Protocol Operators Manage the operational aspects of DeFi platforms, overseeing transactions and user interactions. Liquidity Providers (LPs) Supply assets to liquidity pools, earning fees in return.

Borrowers and lenders engage in lending and borrowing activities, utilizing smart contracts for transactions. Traders execute trades on decentralized exchanges, contributing to market liquidity, and Investors and Speculators Invest in DeFi tokens and projects, often seeking profit from price fluctuations.[15]

Governance participants hold governance tokens, influencing protocol decisions and upgrades. Oracles provide external data to smart contracts, enabling accurate price feeds. Auditors and Security Firms Assess the security of DeFi protocols, identifying vulnerabilities. Decentralized autonomous organizations (DAOs) Facilitate community governance and decision-making. End users utilize DeFi services for personal finance needs.[17]

Insurance providers offer coverage against smart contract failures and other risks. Regulators monitor and enforce compliance within the DeFi space.[25]

While the roles of these participants are generally well-defined, they can evolve over time, particularly as governance structures shift and new technologies emerge [26]. This dynamic nature of participation highlights the complexity and adaptability of the DeFi ecosystem. However, the rapid growth and innovation in DeFi also raise concerns regarding regulatory oversight and the potential for market manipulation, which could impact the entire system's stability.[27]

TABLE I – MAIN ROLES IN DEFI

| Role | Description | Examples | Incentives/ Impact | Challenges |
|---|---|---|---|---|
| Developers | Create and maintain smart contracts and build decentralized applications (dApps) on blockchain platforms. | Developers behind many projects. Some of them are open-source projects. | Drive innovation and introduce new financial products and services. | Ensuring security and scalability of code. |
| Protocol Operators | Maintain and improve the underlying DeFi protocols. | Teams behind Uniswap, Aave, Compound | Ensure the smooth operation of DeFi protocols, impacting liquidity and transaction volume. | Maintaining decentralization while scaling. |
| Liquidity Providers (LPs) | Supply tokens to liquidity pools in decentralized exchanges (DEXs) or lending platforms. | User depositing ETH and DAI in Uniswap pool | Earn transaction fees, interest, or rewards for providing liquidity. | Risk of impermanent loss and market volatility. |
| Borrowers | Use DeFi platforms to borrow cryptocurrencies by providing collateral. | User locking ETH in Aave to borrow DAI | Access funds without intermediaries, possibly lower interest rates. | Managing collateral risks and interest rates. |
| Lenders | Supply assets to lending platforms to earn interest. | User depositing USDC in Compound | Earn interest or rewards for lending assets. | Risk of default, liquidity risks. |
| Traders | Use DEXs to buy, sell, or swap tokens. | Arbitrage traders are exploiting price differences across platforms. | Trade without intermediaries, often at lower costs, and exploit market inefficiencies. | Market volatility, slippage, transaction fees. |
| Investors and Speculators | Invest in DeFi tokens or participate in yield farming to maximize returns. | Users staking tokens in a protocol to earn governance tokens. | Maximize returns through speculation, capital appreciation, or rewards. | Market risk, high volatility. |
| Governance Participants | Token holders who vote on protocol decisions, upgrades, or fee structures. | UNI token holders vote on Uniswap proposals. | Influence protocol development, shaping the future of DeFi protocols. | Low voter participation, governance centralization. |
| Oracles | Provide off-chain data (e.g., price feeds) to DeFi smart contracts. | Chainlink, Band Protocol | Ensure accurate data for functions like lending rates and liquidation triggers. | Ensuring data accuracy and reliability. |

| Role | Description | Examples | Incentives/ Impact | Challenges |
|---|---|---|---|---|
| Auditors and Security Firms | Audit smart contracts for vulnerabilities to enhance trust in DeFi protocols. | CertiK, OpenZeppelin | Enhance the trust and security of DeFi protocols through thorough audits. | Dealing with evolving security threats. |
| Decentralized Autonomous Organizations (DAOs) | Manage and govern DeFi protocols collectively. | MakerDAO managing DAI issuance and collateralization. | Control and manage DeFi protocols in a decentralized manner, ensuring fair governance. | Coordinating consensus among decentralized participants. |
| End Users | Engage with DeFi platforms for various purposes such as trading, saving, or borrowing. | User using a DEX to exchange tokens. | Access financial services without intermediaries, increase financial sovereignty. | Understanding risks, platform usability. |
| Insurance Providers | Offer coverage for risks like smart contract failure or hacks. | Nexus Mutual, Cover Protocol | Protect users against losses from smart contract vulnerabilities or hacks. | Underwriting risks, pricing models. |
| Regulators | Monitor the DeFi ecosystem to ensure compliance with laws and regulations. | - | Ensure DeFi platforms comply with existing laws to protect consumers and the financial system. | Managing decentralized, pseudonymous entities. |

## VI. DEFI ECOSYSTEM: NETWORK CONNECTING THE ACTORS

The participants in DeFi are interconnected in a complex ecosystem, each playing a crucial role in the functioning of decentralized financial services.

Developers create the smart contracts and protocols that form the foundation of DeFi platforms. These protocols are then operated and maintained by protocol operators, who may be part of decentralized autonomous organizations (DAOs). DAOs often govern the protocols through token-based voting systems.[17]

Liquidity Providers (LPs) supply assets to DeFi protocols, enabling various financial services. They interact closely with traders using these liquidity pools to exchange assets. Investors and speculators also participate in these markets, often driven by the search for yield or speculative opportunities.[11]

Lenders and borrowers interact through DeFi lending protocols. Lenders deposit funds to earn interest, while Borrowers use these funds, typically providing over-collateralization due to the absence of traditional credit assessments.[15]

Governance participants, often holding governance tokens, make decisions about protocol parameters and upgrades. Oracles provide external data to smart contracts, which is crucial for many DeFi applications. Auditors and security firms assess the security and reliability of smart contracts and protocols.[28]

End users interact with DeFi protocols for various financial services. Insurance Providers offer coverage against smart contract failures or hacks. Currently having limited involvement, regulators are increasingly scrutinizing DeFi activities to address potential risks.[25]

TABLE II – MAIN ROLES RELATIONSHIP IN DEFI

| Role 1 | Role 2 | Relationship |
|---|---|---|
| Developers | Protocol Operators | Build Protocols |
| Protocol Operators | Liquidity Providers | Attract Liquidity |
| Liquidity Providers | Traders | Provide Liquidity |
| Traders | Protocol Operators | Perform Trades |
| Borrowers | Lenders | Borrow Funds |
| Borrowers | Protocol Operators | Use Lending Platforms |
| Lenders | Protocol Operators | Provide Capital |
| Traders | Oracles | Request Data |
| DAOs | Governance Participants | Governance Decisions |
| Governance Participants | DAOs | Vote |
| Governance Participants | Protocol Operators | Influence Protocols |
| Governance Participants | Developers | Approve Changes |
| DAOs | Protocol Operators | Manage Operations |
| DAOs | End Users | Enable Proposals |
| Investors/Speculators | Protocol Operators | Invest in Tokens |
| Investors/Speculators | Liquidity Providers | Provide Funds |
| Developers | Auditors/Security Firms | Request Audits |
| Insurance Providers | End Users | Offer Insurance |
| Regulators | Protocol Operators | Monitor Compliance |
| End Users | Traders | Interact with DEXs |
| End Users | Borrowers | Borrow Assets |
| End Users | Lenders | Lend Assets |

This table illustrates the key relationships between participants in the DeFi ecosystem, highlighting their respective roles and interactions, being also represented by a network in the following figure.

.

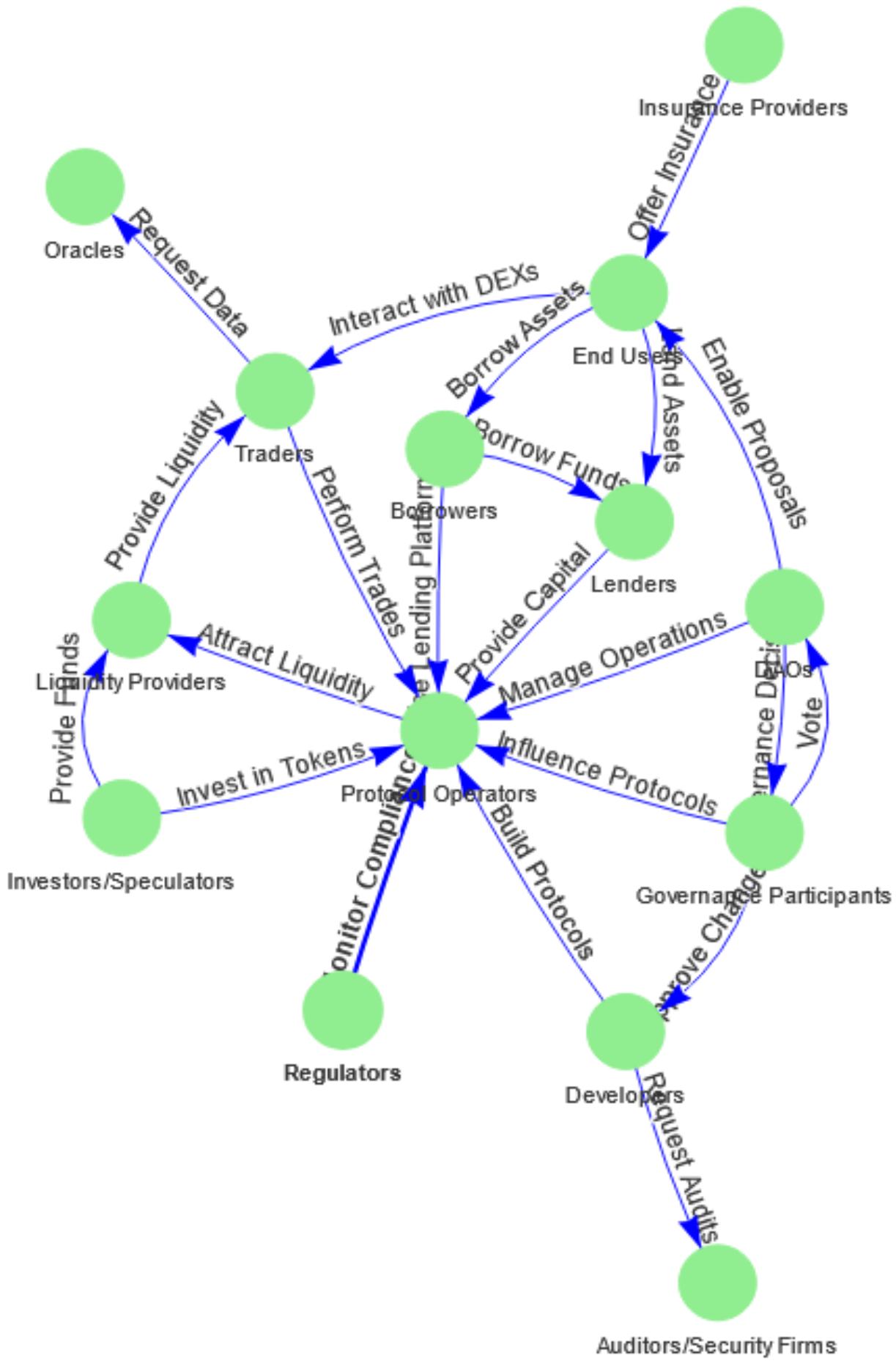

Fig.5. The relationships between roles

## VII. Conclusion

This study underscores the transformative potential of decentralized finance (DeFi) to reshape the global financial ecosystem by offering decentralized alternatives to traditional financial services. The bibliometric analysis highlights a shift in research focus from early technological innovations to growing concerns surrounding sustainability, environmental impact, and regulatory challenges. As the DeFi ecosystem matures, these evolving research priorities reflect a more holistic understanding of the complexities within the space.

The analysis also emphasizes the critical roles played by key participants, including developers, liquidity providers, auditors, and regulators, in shaping the DeFi landscape. Despite the opportunities for enhancing financial inclusion and transparency, the study identifies significant risks that need to be addressed, including smart contract vulnerabilities, liquidity constraints, and regulatory uncertainties.

For DeFi to achieve long-term stability and integration into the broader financial system, stakeholders need to collaborate to foster robust security measures, transparent governance, and effective regulatory oversight. By doing so, the DeFi ecosystem can evolve to offer more inclusive, efficient, and secure financial services. DeFi's participants' interconnected nature and inherent risks call for ongoing innovation and cooperation to ensure its sustainable development and wider adoption.